# Magic Doping and Robust Superconductivity in Monolayer FeSe on Titanates

*Tao Jia, Zhuoyu Chen\*, Slavko N. Rebec, Makoto Hashimoto, Donghui Lu, Thomas P. Devereaux, Dung-Hai Lee, Robert G. Moore, and Zhi-Xun Shen\**

T. Jia[+], Dr. Z. Chen[+], S. N. Rebec, Prof. T. P. Devereaux, Prof. Z.-X. Shen
Stanford Institute for Materials and Energy Sciences, SLAC National Accelerator Laboratory, Menlo Park, California 94025, USA
Departments of Physics, Applied Physics, and Materials Science and Engineering, Geballe Laboratory for Advanced Materials, Stanford University, Stanford, California 94305, USA
E-mail: zychen@stanford.edu, zxshen@stanford.edu

Dr. M. Hashimoto, Dr. D. Lu
Stanford Synchrotron Radiation Lightsource, SLAC National Accelerator Laboratory, Menlo Park, California 94025, USA

Prof. D.-H. Lee
Department of Physics, University of California at Berkeley, Berkeley, California 94720, USA
Materials Sciences Division, Lawrence Berkeley National Laboratory, Berkeley, California 94720, USA

Dr. R. G. Moore
Materials Science and Technology Division, Oak Ridge National Laboratory, Oak Ridge, Tennessee 37831, USA

[+] These authors contributed equally to this work.

**Abstract:** The enhanced superconductivity in monolayer FeSe on titanates opens a fascinating pathway towards the rational design of high-temperature superconductors. Utilizing the state-of-the-art oxide plus chalcogenide molecular beam epitaxy systems *in situ* connected to a synchrotron angle-resolved photoemission spectroscope, epitaxial $LaTiO_3$ layers with varied atomic thicknesses are inserted between monolayer FeSe and $SrTiO_3$, for systematic modulation of interfacial chemical potential. With the dramatic increase of electron accumulation at the $LaTiO_3$/$SrTiO_3$ surface, providing a substantial surge of work function mismatch across the FeSe/oxide interface, the charge transfer and the superconducting gap in the monolayer FeSe are found to remain markedly robust. This unexpected finding indicate the existence of an intrinsically anchored "magic" doping within the monolayer FeSe systems.

The quest for raising superconducting transition temperature ($T_C$) has been a central theme of material science research.[1] A remarkable triumph is the monolayer FeSe grown on SrTiO$_3$ (noted hereafter as 1UC FeSe/STO, UC standing for unit cell), in which superconductivity is significantly enhanced compared to its bulk form.[2-4] Experimental evidence so far suggests that the source of elevated $T_C$ is two-fold: extra electron doping and interfacial mode coupling.[6–9] The role of interfacial coupling effect has been extensively discussed in the literature.[6, 8, 10–15]

To study the doping effect, researchers have employed alkaline metal (Li, Na, K and Cs) atom adsorption for bulk and multilayer FeSe. Due to the low ionization energy alkali metal dosing or intercalation acts as a charge injector. Doping level can be tuned and phase diagrams of $T_C$ are obtained.[9, 16–19] Comparing these phase diagrams, two important pieces of information can be drawn. First, there exists a superconductor-insulator transition in a higher doping regime.[9, 19] Remarkably, transport measurements with Li intercalation exhibit apparent phase-separation features across this transition, indicative of a first-order phase transition instead of a continuous one. Second, a sharp discreteness of possible $T_C$'s is found in the continuous tuning of Li doping, and the discreteness is smoothed with additional disorders, such as replacing Fe with Cu, S with Se, or increasing the size of dopant atoms (e.g. from Li to Na). Although the spatially average carrier concentration is fixed by such doping method, in the presence of disorder, the carrier density distribution is often inhomogeneous. These observations call for a doping method for FeSe without the introduction of impurity dopants.

Monolayer FeSe grown on STO substrate is doped by interfacial charge transfer.[13, 21] Interestingly, the majority of reported high-quality monolayer FeSe films grown on different types of perovskite titanate substrates, including STO(100),[6, 21] STO(110),[14] and BaTiO$_3$[10] with varied dielectric and work function properties, exhibit similar levels of doping (∼ 0.10-0.12 electrons per Fe atom). For increasing the doping ability of the substrate, the LaTiO$_3$ (LTO)/STO heterostructure is an excellent candidate. LTO provides Ti$^{3+}$ and forms a two-dimensional electron gas (2DEG) accumulated at the surface of the LTO/STO heterostructure,[22-24] indicating a lowered surface work function. When growing 1UC FeSe on top of LTO/STO, we would be able to provide an additional chemical potential difference across the interface for charge transfer to FeSe from LTO/STO, while maintaining other properties similar to 1UC FeSe/STO without introducing additional disorders due to the structural similarity between STO and LTO. Different from testing different perovskite substrates for FeSe arbitrarily, we can precisely control the thickness for LTO and thus provide the essential systematics. The *in situ* synchrotron angle-resolved photoemission spectroscope (ARPES) can unambiguously determine the doping of FeSe by Fermi surface volume. Avoiding the use of Li or other alkaline atom adsorption, we can rule out any possibility of Li ordering. Importantly, the insertion of LTO systematically controls the interfacial work function difference for charge transfer, which is thermodynamically distinct from alkaline metal dosing or intercalation, where the number of electrons injected is proportional to the number of adsorption atom number.

In this work, we systematically synthesize 1UC FeSe films on LTO/STO heterostructures in two separate but *in situ* connected molecular beam epitaxy (MBE) chambers, then examine the low-temperature electronic structure of the grown films by *in situ* ARPES in Stanford Synchrotron Radiation Lightsource (SSRL). By varying LTO thickness, we find that the itinerant electron density at the surface of the LTO/STO heterostructure surges to more than 4 × 10$^{14}$ cm$^{−2}$, but surprisingly neither the doping nor the superconducting gap of 1UC FeSe film grown on it exhibit noticeable changes. Our results show that the superconductivity in 1UC FeSe thin films is robust and accompanied with an anchored "magic" doping level.

We grow LTO films on STO substrates after the growth of STO buffer layers in a shutter-controlled oxide MBE chamber, monitored by *in situ* Reflective High Energy Electron Diffraction (RHEED). The STO buffer layer is grown using a shuttered approach for deposition of different elements.[25–27] To grow LTO, we use a shuttered approach with on-the-fly adjustment of the shutter times layer-by-layer to maximize the RHEED intensity oscillations. We then transfer the LTO films *in situ* to a separate chalcogenide MBE chamber for the growth of 1UC FeSe. After vacuum post-annealing, the samples are transferred *in situ* to the ARPES chamber at SSRL beamline 5-2 for measurement. We control the annealing conditions identical for different samples to avoid annealing-related variation.[28,29] More details of growth and measurement conditions can be found in Supporting Information.

The ARPES spectra of a 1UC FeSe/5UC LTO/STO heterostructure sample are shown in **Figure 1** as a representative example. It has a Fermi surface with only electron pockets near M point (Brillouin zone corner), and a Luttinger volume count that gives $0.11 \pm 0.01$ electrons per Fe atom (Figure 1b). The top of the hole bands at $\Gamma$ are $\sim 75$ meV below Fermi level, and the bottom of electron bands at M is about 55 meV below Fermi level (Figure 1a and 1c). Replicas of electron and hole bands are also visible in the spectra at M point, which is clearer after taking 2nd energy derivative of the image (Figure 1d). At low temperature, there is a superconducting gap of 14 meV with clear back bending, as is shown in the Energy Distribution Curves (EDCs) in Figure 1e. The gap basically disappears at 57 K or above (Figure 1f). The band structure features of 1UC FeSe on 5UC LTO film resemble that of 1UC

FeSe/STO, which is unexpected considering the additional electrons provided by La in the LTO/STO.

To probe this artificial structure more systematically, we measure 1UC FeSe grown on LTO films with different thickness, as shown in **Figure 2**. All the 1UC FeSe films have almost identical doping level regardless of whether it is on STO substrate or any thickness of LTO films (Figure 2a – 2d). In sharp contrast, the electronic structure changes dramatically from STO to LTO with different thickness prior to the FeSe deposition (Figure 2e – 2h). The STO substrate (pre-annealed with the same condition for later FeSe growth) shows clear Fermi surfaces, consistent with previous studies.[30, 31] The three largest Fermi surfaces consist of one circular $d_{xy}$ subband, one horizontally elongated oval $d_{yz}$ subband, and one vertically elongated oval $d_{xz}$ subband. There exist Fermi surfaces originated from higher-order subbands, but since they are much smaller and the electrons occupying these subbands are much farther away from the surface, these higher order subbands are less relevant to our focus. When we grow 1UC LTO on top of STO, all the three Fermi surfaces become significantly larger.[24] As the number of LTO layers increase, more electrons are provided by La, giving rise to even larger Fermi surfaces. For the 3UC and 20UC cases shown in Figure 2g and 2h, the edges of the Fermi surfaces of the $d_{yz}$ and $d_{xz}$ subbands extends beyond the Brillouin zone. This observed surge of accumulated electrons at the surface of the LTO/STO structure is a result of the deepened confinement potential well at the LTO/STO surface, indicative of a decreased surface work function. Note that minor portion of measured electrons are possibly associated with oxygen vacancies induced in a double Auger process with photon exposure to 84 eV (higher than 38 eV) photons until saturation,[32, 33] yet the electrons measured in LTO/STO exceed that of STO substrate by far even before such exposure to high-energy photons, as shown in Supporting Information.

We summarize our results in **Figure 3**. Figure 3a shows the density of the accumulated electrons at the surface of LTO/STO with different LTO thickness. We count the lowest $d_{xy}$, $d_{yz}$ and $d_{xz}$ bands in the ARPES Fermi surface maps (higher-order subbands are much smaller and the electrons are located much deeper away from the interface, thus they are much less relevant to the interfacial effects we focus on). Electron density quickly increases as the LTO film thickness changes from 0UC (bare STO) to 3UC, and reaches a plateau of $\sim 5 \times 10^{14}$ cm$^{-2}$ for 3UC and thicker LTO films. The saturation behavior is consistent with numerical simulation in Supporting Information. Electron density observed from the surface-sensitive ARPES for thinner LTO films are lower, due to the electron redistribution between LTO and the STO layers. LTO films have effectively lower work functions, thus certain amount of itinerant electrons will be transferred to the STO layers. When LTO films are thicker, the interface between LTO and STO, where the charge transfer occurs, becomes deeper and less influential to the surface, resulting in an increased and saturated observable electron density. Based on the simulation in the Supporting Information, with the dense electron accumulation for LTO thicker than 3UC, the surface work function of the LTO/STO is lowered by $\sim 0.7$ V.

In sharp contrast, the electron density as measured by ARPES Fermi surface maps and superconducting gap at low temperature for 1UC FeSe/LTO/STO films remain nearly unchanged with different LTO thickness, as plotted in Figure 3b. For 0UC (STO bare substrate), the 1UC FeSe film on top exhibits higher doping than the STO substrate prior to deposition of FeSe. Starting from 1UC LTO, the electron density on the surface of LTO/STO is larger than that of the 1UC FeSe grown on top. For the cases where the LTO thickness is 3UC and above, the electron density of LTO/STO is about 3 times that of the 1UC FeSe. Regardless of the dramatic changes of the electron density in the LTO/STO substrate, both electron density and superconducting gap of 1UC FeSe remain basically the same within experimental errors. The doping of 1UC FeSe/LTO/STO falls largely between 0.10 and 0.12 electrons per Fe atom, or between $1.31 \times 10^{14}$ and $1.58 \times 10^{14}$ cm$^{-2}$, and the gap is mostly between 12 meV to 16 meV.

As the doping level is anchored for 1UC FeSe despite large changes in the substrate, an immediate implication could be that the doping is from the monolayer FeSe film itself alone, such as Se vacancies.[34, 35] Yet, this requires a blockage of electron tunneling between FeSe and the oxide substrate so that the work function across the interface does not need to be balanced. This is highly unlikely since it contradicts direct experimental evidence of charge transfer.[13, 20] As long as such blockage is not established, interface details between epitaxial 1UC FeSe and the underlying oxide do not change the nature of the metal-metal contact. As electrons move across the interface between FeSe and the titanate substrate, an energy equilibrium state would be established by the redistribution of electrons. Here, we increase the electron density in titanate substrates to $\sim 5 \times 10^{14}$ cm$^{-2}$, much larger than the typical electron density of monolayer FeSe/STO, not only effectively creating metal-metal contact, but also greatly lowered the effective work function of the substrate surface. After FeSe is grown on LTO/STO, for balancing additional work function difference across the interface, FeSe electron density would be expected to be much higher than the case on STO substrates (see more details in Supporting Information). Therefore, the ARPES measured unchanged doping of 1UC FeSe/LTO/STO points to some unusual intrinsic properties of FeSe.

Similar phenomenon of robust superconductivity is also seen in the Li-intercalated FeSe thin flake experiment,[19] where discrete $T_C$ changes are observed as Li is continuously intercalated (see Figure 3c). After the system reaches and plateaus at the highest $T_C$ (~ 44 K), further doping brings the system into an insulator with marked discreteness in transport measurements. The superconductor-insulator phase transition is also observed in K dosing experiments on multilayer FeSe thin films.[9] The doping for maximal $T_C$ (~ 45 K) is ~ 0.11 electrons/Fe atom, similar to the case of 1UC FeSe on STO or LTO. Above this doping, the system gradually transitions into an insulating phase. The major difference between the Li and K experiments is the level of discreetness. The K experiment seems to be a smoothed version of the Li experiment. In our work, we find an unusually anchored doping level despite strong interface electron accumulation as a "clean" doping channel, since we do not introduce extra disorder by ad-atoms or vacancies. Moreover, as a method of varying interfaical work function difference rather than direct injection of electrons (as in Li or K experiments), by LTO insertion we have not observed obvious signatures of the insulating phase nor phase separation. These results are summarized in Figure 3c, where we combine our observations on monolayer FeSe/LTO/STO, in which the doping is anchored at 0.11 ± 0.01, with the Li and K experiments.[9, 19]

Our results of a "magic" anchored doping on 1UC FeSe despite large variation in substrate carrier density, combined with the studies listed above, show that the doping of FeSe is far from being fully understood. Below we propose a scenario that could possibly explain the unique phenomena of FeSe doping levels. In theory, there might exist a first-order phase transition between a superconducting phase with a maximum possible doping of ~ 0.11 and an insulating phase at a higher doping governed by a yet concealed order for FeSe. In real material, true first-order phase transition never exists with the presence of disorder, but by approaching the "clean" limit, first-order-like behaviors, such as phase separations, could be observed. Thus, electron injection higher than ~ 0.11 by Li dosing to FeSe could have formed phase separations between the superconducting phase and the insulating phase.[19]

However, in the FeSe/LTO/STO system, the extra work function difference, which would facilitate higher electron transfer at the interface in a trivial case, need to exceed a critical potential barrier originated from the theoretical first-order phase transition to transit into the insulating phase. Our observation of an anchored doping and absence of the insulating features suggests such critical potential barrier is still higher than the increased work function difference built by LTO insertion. Even if small amount of insulating phase would exist and form phase separations due to finite temperature, only the ~ 0.11 doping superconducting phase would be visible by ARPES, and the minor portion of insulating phase would be hard to discern by APRES due to the low intensity and the defuse nature of the spectra.[9] This scenario might also explain why most reported high-quality monolayer FeSe films grown on different types of perovskite titanate substrates[6, 10, 14, 21] "magically" exhibit similar ~ 0.11 doping in ARPES, which coincidentally corresponds to the maximum $T_C$ found in doped multilayer systems.[9, 18] In the cases of K/Cs dosing or excess Se, additional disorders are introduced and the discreteness is smoothed, making the transitions more continuous and less first-order-like.[36] This leads to a continuous change in Fermi surface volume and $T_C$.[7, 9, 18, 37, 38] Interestingly, the $T_C$ evolution with Na intercalation represents an intermediately smoothed case between Li and K.[19]

Another aspect of our results is the largely invariant superconducting gap under great change of itinerant electron density in the substrate. For the interfacial electron-phonon coupling with phonon modes in directions parallel to the surface, dense itinerant 2DEG on LTO surface can provide strong screening effect. However, because the in-plane motion of carriers in the substrate cannot screen the charge transfer induced electric field that is perpendicular to the interface, we expect the extra carriers in LTO cannot screen the long-wavelength longitudinal optical phonon modes associated with ionic vibrations that are also perpendicular to the interface and modulating the interfacial electric field. It is precisely this type of phonon which is suggested to enhance the superconductivity in 1UC FeSe/STO system.[6, 39, 40] Therefore, the largely unchanged superconducting gap of FeSe on LTO with different thicknesses suggests that the relevant interfacial coupling at FeSe/oxide interface are strongly selective for the phonon modes involved.

In conclusion, empowered by the *in situ* oxide and chalcogenide MBE systems that are directly coupled to the synchrotron ARPES, we have systematically studied the electronic structure 1UC FeSe on LTO/STO heterostructures. We find that the doping level and enhanced superconductivity of the monolayer FeSe is exceptionally robust in spite of substantial increase of electron density in the substrate. The indicated anchored "magic" doping level suggests a unique underlying material property, posing a challenging target for theoretical and computational materials science research.

**Acknowledgements**
T. Jia and Z. Chen contributed equally to this work.
We thank B. Moritz and C. D. Pemmaraju for helpful discussions. This work is supported by the Department of Energy, Office of Science, Basic Energy Sciences, Materials Sciences and Engineering Division, under Contract DE- AC02-76SF00515. Use of the Stanford Synchrotron Radiation Lightsource, SLAC National Accelerator Laboratory, is supported by the U.S. Department of Energy, Office of Science, Office of Basic Energy Sciences, also under Contract No. DE-AC02-76SF00515. D.H.L. was funded by the U.S. Department of Energy, Office of Science, Office of Basic Energy Sciences, Materials Sciences and Engineering Division under Contract No. DE-AC02-05-CH11231 within the Quantum Materials Program (KC2202). R.G.M. is supported by the Laboratory Directed Research and Development Program of Oak Ridge National Laboratory, managed by UT-Battelle, LLC, for the U. S. Department of Energy.

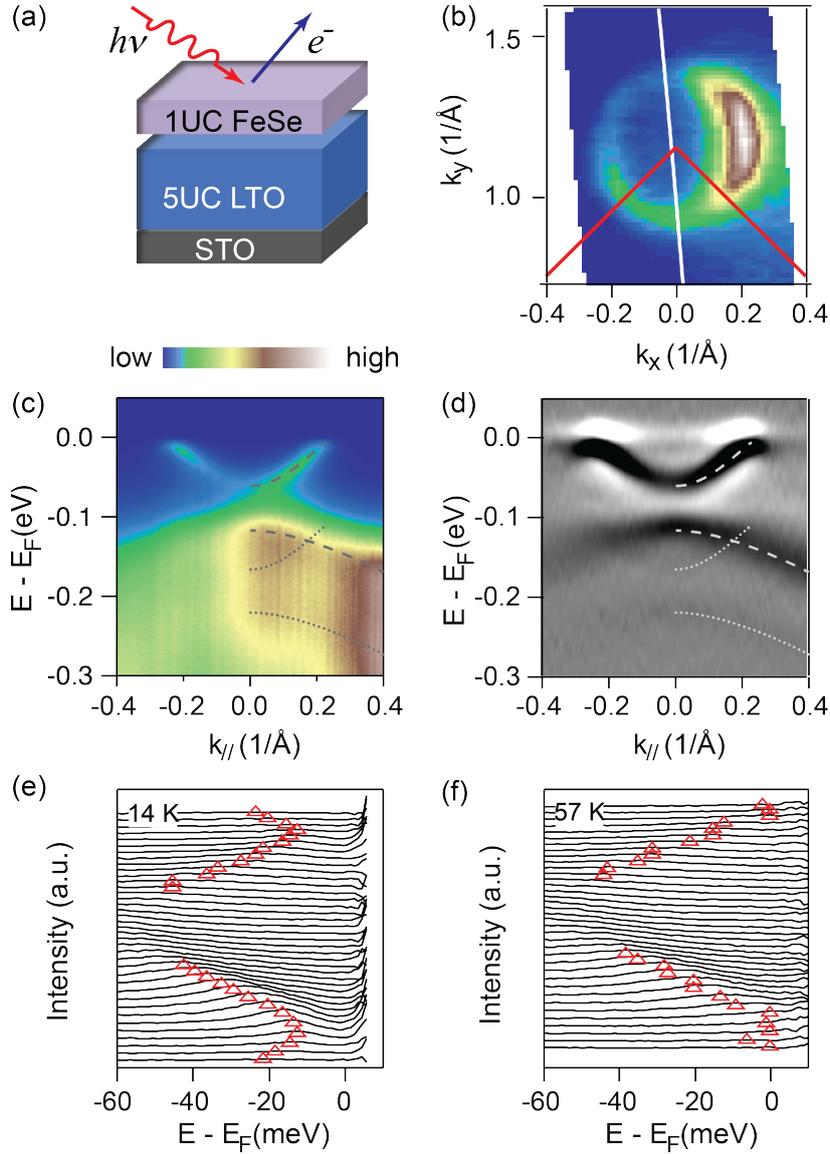

**Figure 1.** ARPES characterizations of 1UC FeSe/5UC LTO/STO films. (a) Schematic diagram of the material structure. (b) The Fermi surfaces of electron pockets near the zone corner M. Red lines indicate the Brillouin zone edges. (c-d) Spectra and its 2nd derivative taken at zone corner M, along the cut shown with the white line in (b). The dashed and dotted curves are guides to the eye for the main bands and the replica bands. (e,f) Energy distribution curves (EDCs) for the spectra taken at the zone corner M. The EDCs are divided by Fermi distribution function at measurement temperatures 14 K and 57 K, respectively. Red triangles indicate the energies of maximum intensities within $E - E_F = [-50\ \text{meV}, 0]$. All spectra here are taken with 28 eV photons.

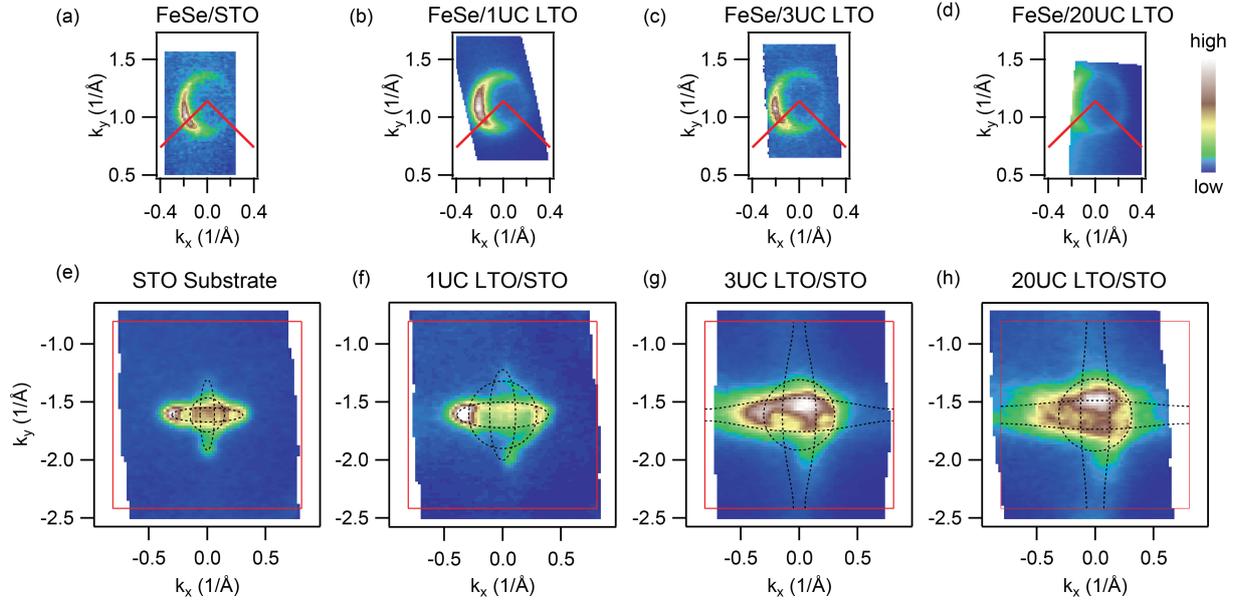

**Figure 2.** Systematic Fermi surface maps of FeSe and LTO/STO heterostructures. (a-d) The Fermi surface maps near M for 1UC FeSe films on STO substrate, 1UC, 3UC, and 20UC LTO films, respectively, taken with photon energies between 25 and 28 eV. (e-h) The Fermi surface maps of STO substrate, 1UC, 3UC, and 20UC LTO films grown on STO, respectively, taken with photon of 84 eV and circular right polarization. Red lines indicate the Brillouin zones. Dashed lines are guides to the eye of the three outermost Fermi surfaces. To estimate the Fermi surface sizes, both maps with circular right and linear vertical (Fig. S3 and S5) polarizations are used.

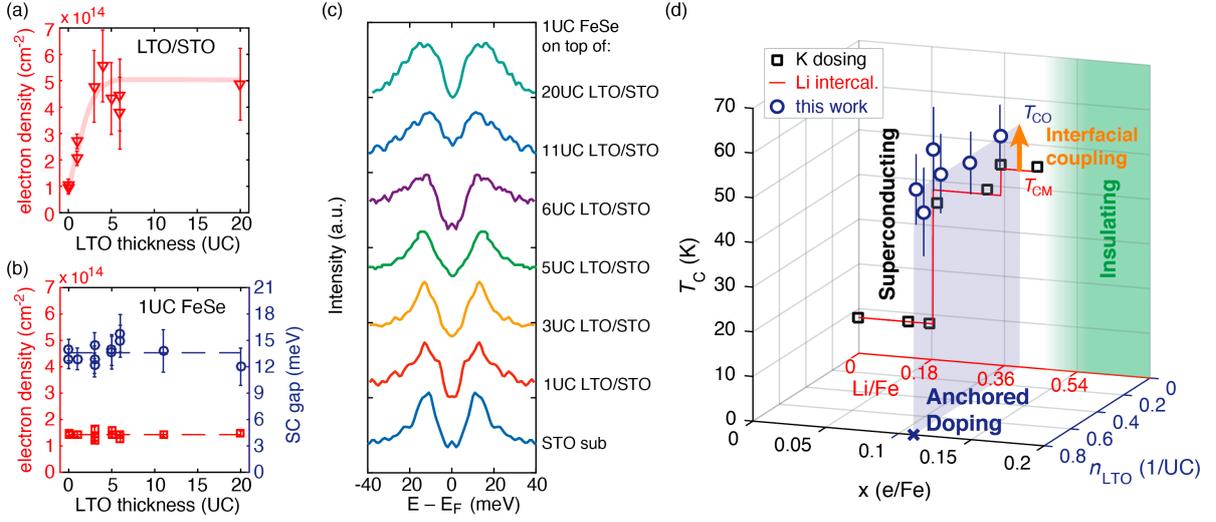

**Figure 3.** (a) ARPES measured electron density of LTO films as a function of film thickness. The red thick curve is a guide to the eye for the electron density trend of increase and saturation. (b) (left) Electron density of 1UC FeSe on LTO/STO heterostructures with different LTO thickness, in red squares. (right) Superconducting gap at temperatures below 20 K for 1UC FeSe on LTO/STO heterostructures with different LTO thickness, in blue circles. Dashed lines show the average values for all samples. (c) Representative symmetrized EDCs at $k_F$ with measurement temperatures lower than 20 K for different thickness of LTO insertion. (d) The blue open circles show the $T_C$ and doping in the unit of electron per Fe atom of 1UC FeSe/LTO/STO with various LTO/STO electron concentration $n_{LTO}$, in the unit of electron per in-plane LTO/STO unit cell, as measured by ARPES Fermi surface maps. $T_C$ values are converted using superconducting gap data, using a coefficient $2\Delta_0/k_B T_C = 5.7$, consistent with literature.[12] The transparent blue vertical plane corresponds to the average values of doping and $T_C$, denoted as $T_{CO}$, in which "O" stands for oxide substrates. The black squares are $T_C$ data extracted from Figure 4 in ref.[9] (a K dosed FeSe ARPES experiment), as a function of electron doping x (e/Fe) measured by ARPES Fermi surface maps. The red curve is a reproduction of discrete $T_C$ steps for lithium ionic solid gated FeSe thin flakes, as a function of nominal Li content with Li/Fe ratio adapted from Figure 5 of ref.[19]. Note that the content (Li/Fe) axis is re-scaled to match the actual electron doping axis x (e/Fe) as measured by ARPES Fermi surface maps. $T_{CM}$ represents the maximum $T_C$ recorded in doped bulk/multilayer FeSe systems. The insulating regime indicated by the green shaded area represents the findings from both references.[9, 19]

# Supporting Information

## Magic Doping and Robust Superconductivity in Monolayer FeSe on Titanates

*Tao Jia, Zhuoyu Chen\*, Slavko N. Rebec, Makoto Hashimoto, Donghui Lu, Thomas P. Devereaux, Dung-Hai Lee, Robert G. Moore, and Zhi-Xun Shen\**

**Section 1. Synthesis of FeSe/LTO/LTO heterostructures and ARPES measurements**

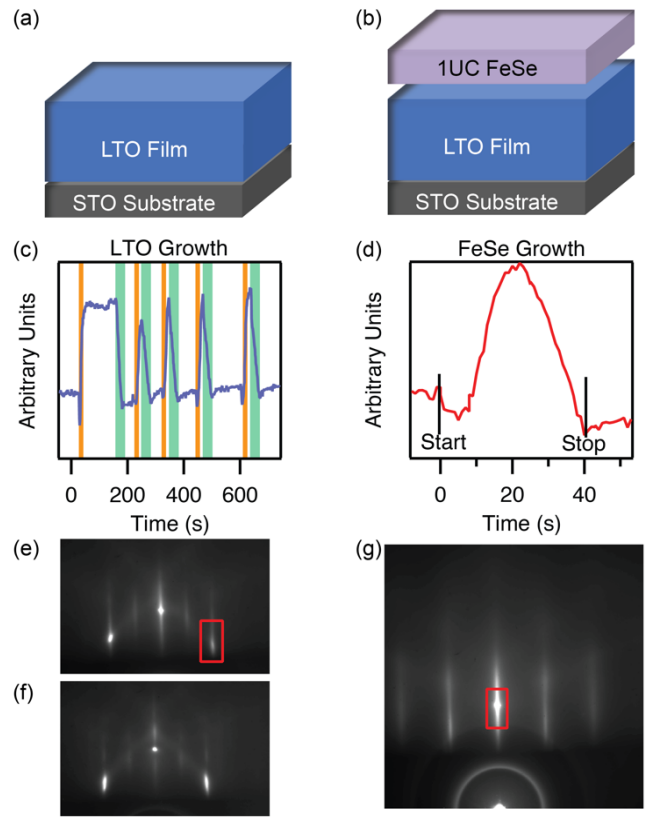

**Figure S1**. Sample growth. (a-b) Schematic diagrams of LTO/STO and 1UC FeSe/LTO/STO heterostructures. (c) The RHEED oscillations of a shutter-controlled 5UC LTO film growth as an example. The vertical axis is the integrated intensity within the red window in (e). Orange and green colored backgrounds indicate the time windows when the La and Ti shutters are opened, respectively. White backgrounds correspond to pauses during the growth. (d) The RHEED oscillations of a 1UC FeSe film on the grown 5UC LTO film shown in (c) and (e). The vertical axis is the integrated intensity within the red window in (g). (e) The RHEED pattern after the growth along the STO [100] direction of a 5UC LTO film. (f) The RHEED pattern after the growth along the STO [100] direction of another 5UC LTO film, for which the shutter time is adjusted from layer to layer. (g) The RHEED pattern after the growth along the STO [100] direction of a monolayer FeSe film on the 5UC LTO film shown in (e).

All samples are grown on 0.05% Nb Doped STEP STO purchased from Shinkosa. The substrates are mounted to inconel sample holders using silver paste. The samples are first exposed to $6\times10^{-6}$ Torr oxygen partial pressure and heated to 900 °C for substrate annealing for 30 minutes before the LTO growth at 820 °C in the same oxygen environment. The RHEED is aligned along the STO [110] direction during growth. Before an LTO growth, an STO buffer layer of 5-20 unit cells (UC) is first grown on the STO substrate to ensure the surface smoothness and consistency. Three cells are used for growth: a differentially pumped source loaded with ultra high purity Sr (99.95%), a high temperature cell loaded with ultra high purity Ti (99.995%), and a high temperature cell loaded with ultra high purity La (99.995%). The source flux and deposition rates are calibrated and set using a quartz crystal microbalance.

Growth is done using a shuttered approach. A typical LTO (STO) recipe starts with the La (Sr) shutter opening and ends with the Ti. Before growth, we usually align the tilting angle of the STO substrate to maximize the central streak intensity.[1] The STO buffer layer is grown using a shuttered approach for deposition of different elements.[2–4] To grow LTO, we use the shuttered approach with on-the-fly adjustment of the shutter times layer-by-layer to maximize the RHEED intensity oscillations. Namely, we don't seek to keep the shutter time exactly consistent from layer to layer for both La and Ti, but with a general shutter time guideline that is found from previous test growths. We majorly focus on maximizing the RHEED quality during the growth. Focusing on the (01) streak intensity oscillation as shown in **Figure S1e**, if we saw a "double-peak" feature (similar to what is described in ref.[2]) in the previous RHEED oscillation, we typically decreased the La shutter time by shutting the shutter slightly before the RHEED reached peak intensity. If we saw an obviously lower peak intensity, we typically increased the La shutter time so that The RHEED intensity went higher. If we saw a lower valley intensity than general, we decreased Ti shutter time by shutting the shutter slightly before the RHEED reached typical valley intensity. If we saw a higher valley intensity than general, we increased Ti shutter time by shutting the shutter only when the RHEED reached typical valley intensity. By this adjustment strategy, the RHEED oscillations generally converge to a largely consistent amplitude. This strategy brings about a shutter time variation. Among samples that we have measured in ARPES (15 films), the standard deviation for La shutter time is 14% (change from average shutter time for each layer) and standard deviation for Ti is 8%. We discuss the effect of the shutter time variation below. Different from the case shown in Figure S1c in which each layer having exact same shutter time, Figure S1f corresponds to a film of same 5UC LTO thickness but with the first layer having ∼ 20% shorter and second layer having ∼ 50% longer La shutter time compared to the average. This longer shutter time is compensated by shorter shutter time in latter layers. This variation in shutter time could be one source of cation disorder since it would require redistribution of cations within ∼ 1 unit cell thickness. We find that, at the relatively high growth temperature (820 °C) we use, the effect of shutter time variation is not obvious in the final RHEED patterns, as can be seen in the comparison between Figure S1e and S1f, indicating the cation disorder level on the surface is lower than what RHEED can resolve. This is consistent with the finding that shown in ref.[4], in which we intentionally change the growth so the oscillations got "flipped" in phase, but after the oscillations got "flipped" back, a good final oscillation can still be obtained and ARPES still shows a surface electron accumulation after UV radiation. We speculate that the surface sticking coefficient gradually changes from STO buffer layer to thicker and thicker LTO films, but further experiments and simulations are needed to clarify. After closing each shutter, the growth is usually paused for RHEED intensity to saturate. After saturation, longer pause does not have an additional noticeable effect on the growth of the film. Post growth the samples are cooled down in oxygen background and then transferred in situ to chalcogenide MBE chamber or ARPES chamber for further growth or measurement.

FeSe growths are done in a chalcogenide MBE chamber, at a base pressure of $8 \times 10^{-11}$ torr. The LTO/STO films are transferred *in situ* to the chalcogenide chamber and heated up to 360 °C for growth without additional annealing. Ultra-high purity iron (99.995%) and selenium (99.999%) were then deposited onto the substrate. The film was then annealed at 500 °C for 2 hours. The growth conditions of FeSe on STO (001) are described in ref.[5], except that after growth they are annealed at 500 °C for 2 hours for consistency. Figure S1d and S1g shows the RHEED oscillation during growth and the resulting RHEED pattern, respectively.

After growth, the films are transferred in situ to the ARPES end station of the Stanford Synchrotron Radiation Lightsource beamline 5-2. The base pressure in the ARPES chamber is lower than $4\times10^{-11}$ Torr. The measurements on FeSe are done with photon energy of 25-28 eV, and the best energy resolution for gap measurement is ~ 6 meV. The measurements on LTO films and STO substrates are done using photon energy of 84 eV, and the energy resolution is better than 31 meV. The angular resolution is better than 0.1º.

## Section 2. ARPES spectra of 1UC FeSe/LTO/STO and LTO/STO

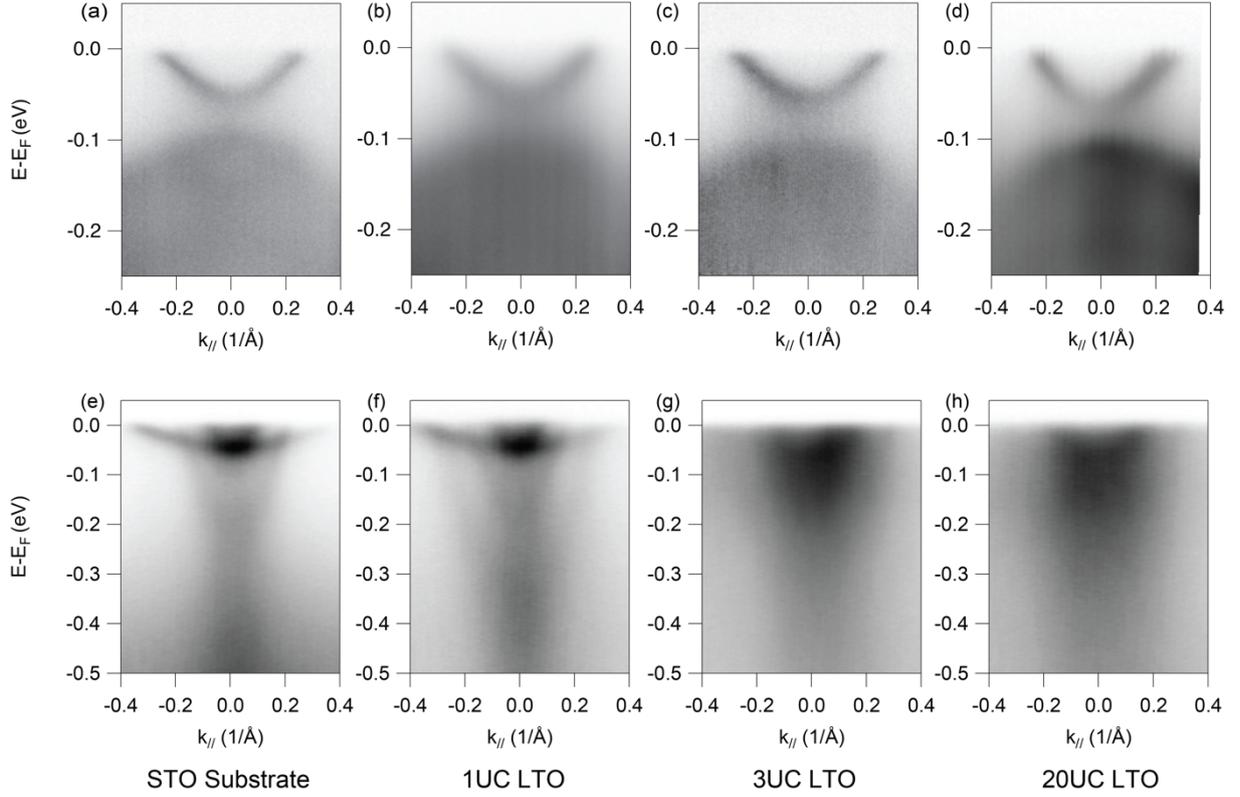

**Figure S2**. ARPES spectra of FeSe/LTO/STO and LTO/STO films with different thickness along high symmetry cuts. (a-d) ARPES spectra of 1UC FeSe/LTO/STO heterostructure with different LTO thickness near zone corner along the direction from zone center to zone corner. The number of LTO layers are 0 (STO substrate), 1, 3, 20, respectively. (e-h) ARPES spectra of LTO/STO heterostructure at the center of the second Brillouin zone $(0, -2\pi/a)$, where a is the lattice constant of the heterostructure. The number of LTO layers are 0 (STO substrate), 1, 3, 20, respectively.

## Section 3. Polarization dependence for LTO/STO ARPES

The Fermi surface map of LTO films displayed in Figure 2 in main text is measured with circular right (CR) polarization. **Figure S3** (e-h) shows that the elliptical pocket along the direction of $k_y$ is missing if linear vertical (LV) polarization is used. This is consistent with previous research on STO 2DEG.[6]

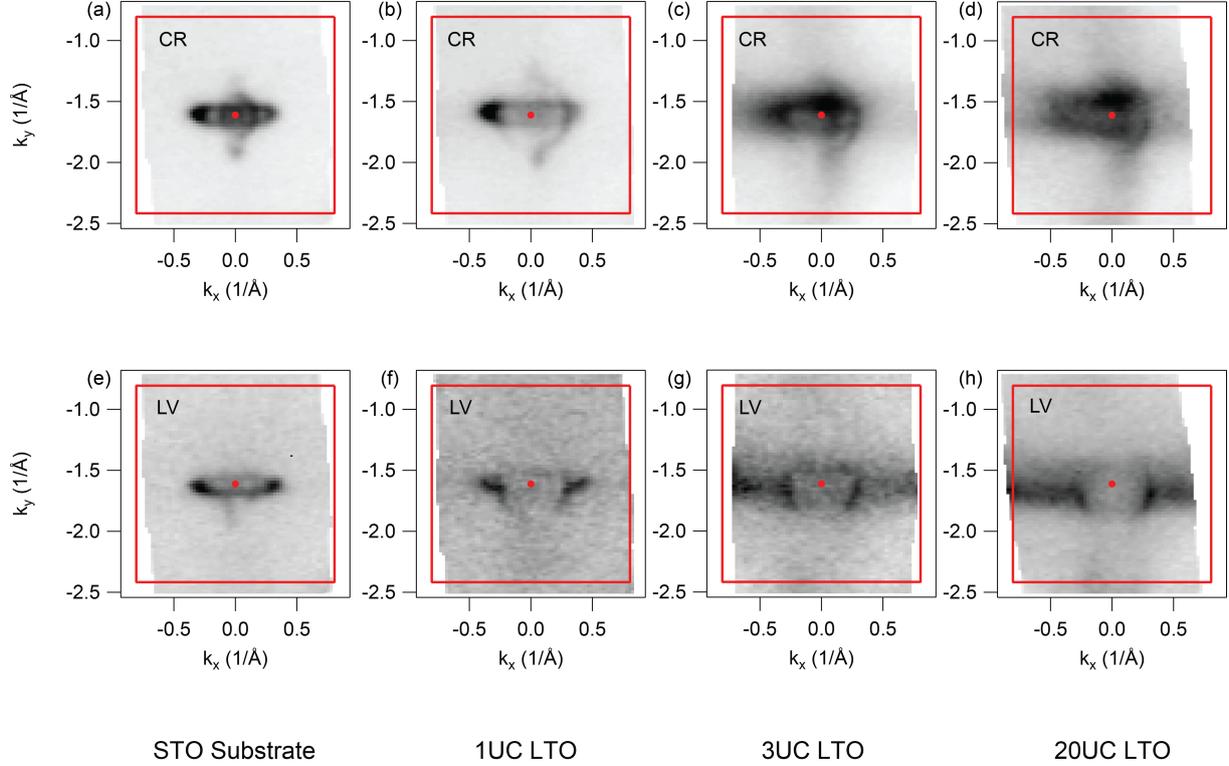

**Figure S3**. (a-d) Fermi surface map of LTO/STO heterostructure with different LTO thickness taken with circular right (CR) polarization. These are the same figures as Figure 2 (e-h) in main text. The number of LTO layers are 0 (STO substrate), 1, 3, 20, respectively. (e-h) Fermi surface map taken at the same setting as (a-d), except with linear vertical (LV) polarization.

## Section 4. ARPES spectra before exposure to high-energy photons

2DEGs can emerge on STO surface after exposure to high-energy photons. This process, as is discussed in the main text, is achieved by oxygen vacancies created in a double-Auger process.[7, 8] However, in order to acquire high-quality spectra of 2DEGs on LTO/STO, one has to measure the centers of higher Brillouin zones, thus a higher photon energy than the Auger process threshold (38 eV) is necessary. In our research, most data on LTO/STO are taken with 84 eV photons. Note that the double Auger effect is a relatively minor effect compared to the amount of electron provided by LTO. Here we discuss this effect for completeness.

To avoid the photon-induced electrons, we measure the first Brillouin zone ($\Gamma(0, 0)$) using 28 eV photons. Although the Fermi surfaces are only partially visible at $\Gamma(0,0)$, likely due to matrix element effects, important conclusions can still be drawn. From **Figure S4**, we can see that without radiation of 84 eV photons the pockets are bigger for 1UC LTO compared to that of STO, and even bigger for 5UC LTO. This demonstrates that there are intrinsic electrons without the exposure to higher energy photons. It also supports our findings in Figure 2 and 3 in the main text that the electron density increases rapidly as LTO thickness increases from 0UC (STO) to 1UC to more than 3UC. Furthermore, this is consistent with our understanding that photon-induced electrons are created in a parallel channel (from oxygen vacancies) to the intrinsic electrons of LTO (from La-donated electrons), thus the electrons from the two processes should be largely additive.

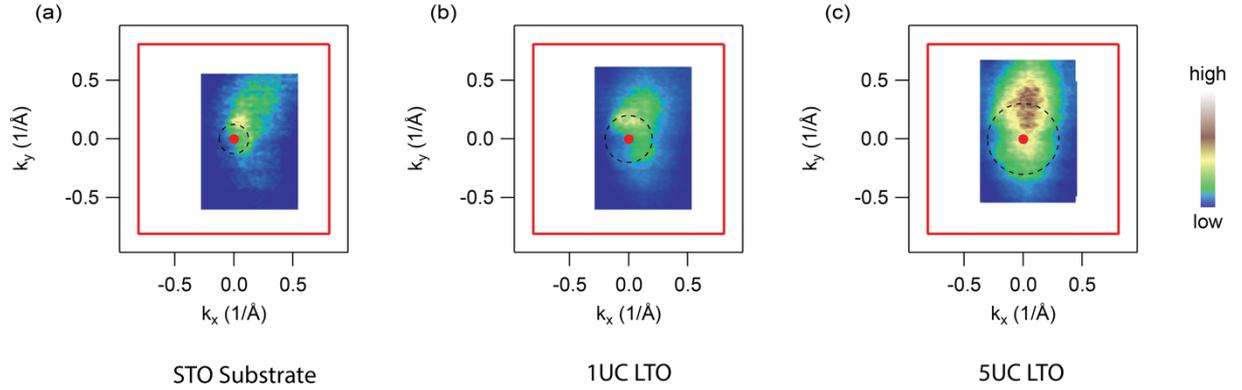

**Figure S4**. LTO/STO heterostructures measured with 28 eV photons. (a) intensities near EF for STO substrates without exposure to 84 eV photons. (b-c) same as top left, but for 1UC and 5UC LTO films on STO substrates, respectively.

## Section 5. Additional data of LTO/STO 2DEG

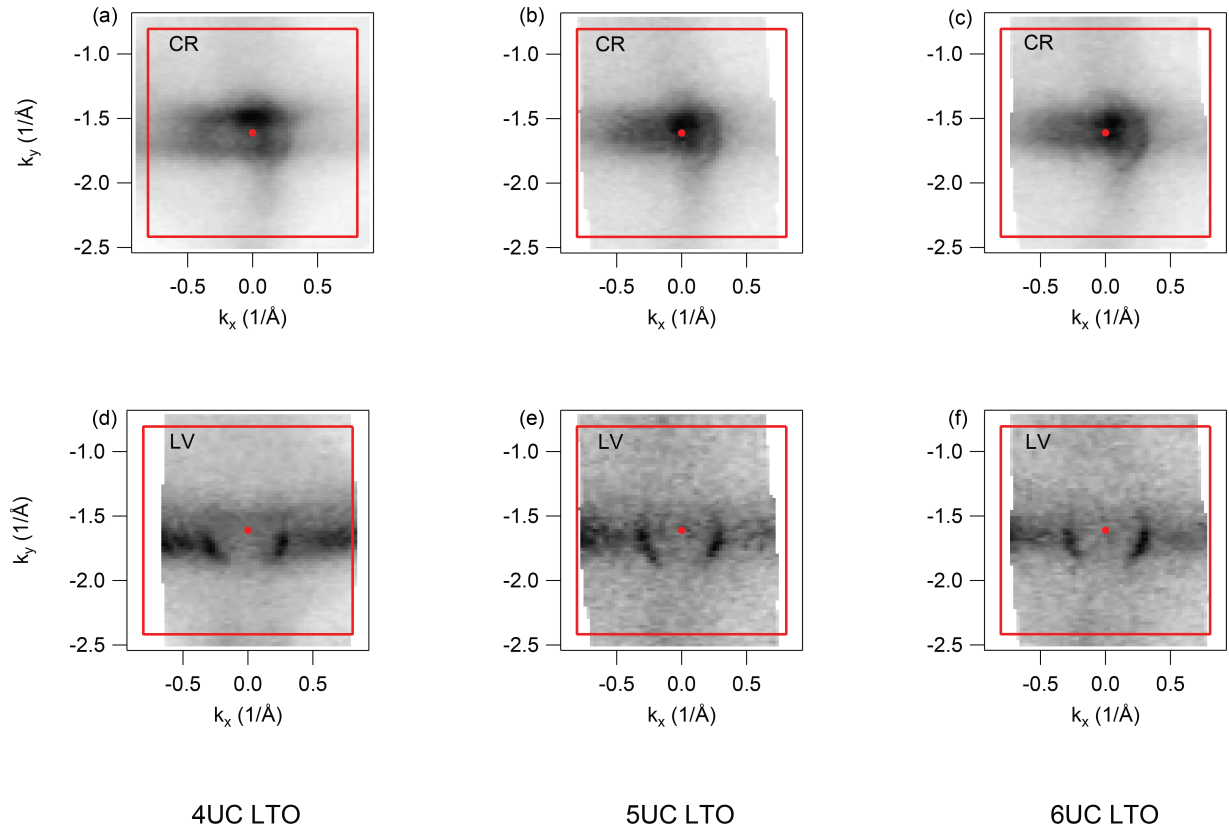

                4UC LTO                   5UC LTO                   6UC LTO

**Figure S5**. (a-c) Fermi surface maps of 4, 5 and 6UC LTO films measured with circular right (CR) polarized photons. (d-f) Fermi surface maps of 4, 5 and 6UC LTO films measured with linear vertical (LV) polarized photons.

# Section 6. Temperature and LTO thickness dependence of superconducting gap for 1UC FeSe/LTO/STO

The temperature dependent superconducting gap was extracted from the energy distribution curves at $k_F$ symmetrized at $E_F$ for multiple samples, yielding similar $T_C$ as 1UC FeSe/STO, which is consistent with the gap size at lowest temperatures. Figure S6 (a-d) show the temperature dependence of superconducting gap for a 1UC FeSe/5UC LTO/STO samples and a 1UC FeSe/11UC LTO/STO sample. The gap values as a function of temperature are fitted using the formula $\Delta(T) = \Delta_0 \tanh(k\,(T_C/T - 1))$. Figure S6 (e) shows the symmetrized EDCs for lowest temperature gap (< 20 K) of FeSe samples with different LTO thickness.

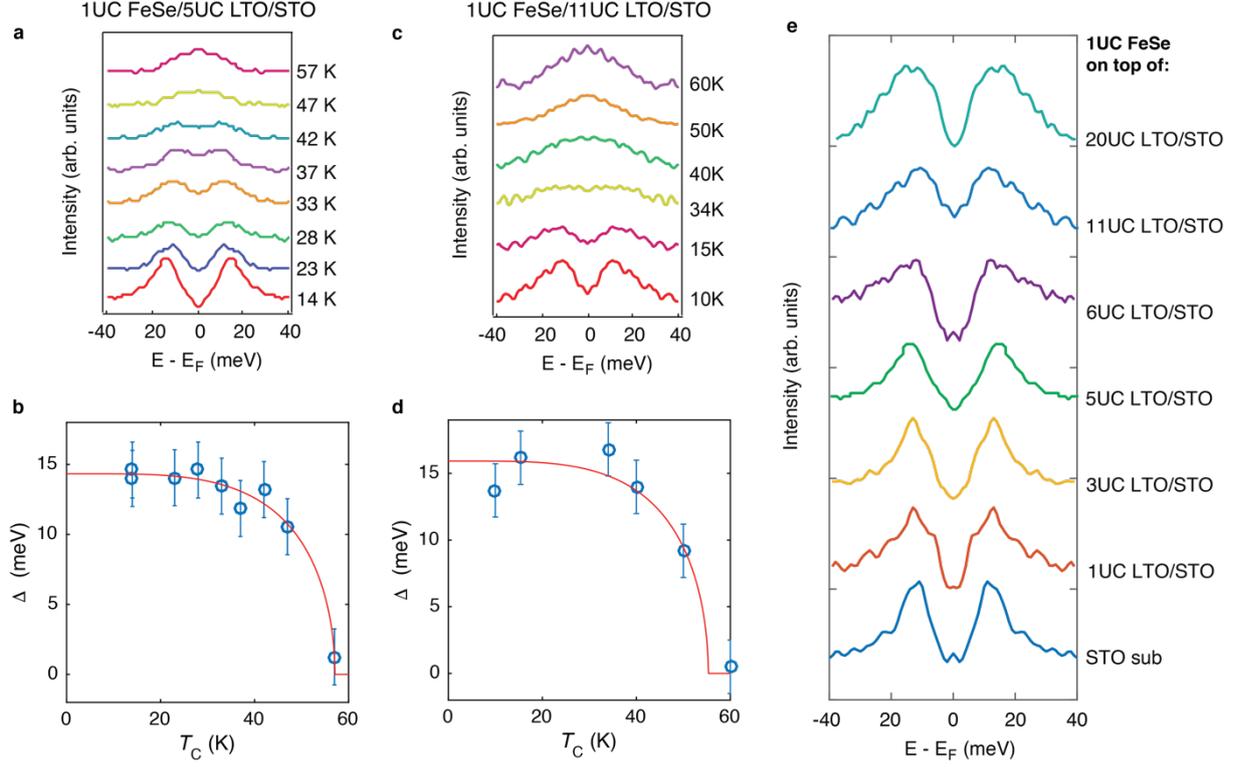

**Figure S6**. Temperature dependence of superconducting gap for a 1UC FeSe/5UC LTO/STO sample (a,b) and a 1UC FeSe/11UC LTO/STO sample (c,d). (a,c) Energy distribution curves (EDCs) at $k_F$ at different temperatures, symmetrized at $E = E_F$. (b,d) Temperature dependence of superconducting gap. Blue open circles are measured gap values extracted using the model in ref.[9], and red curve is a mean-field fitting. The fitting of the gaps yields $\Delta_0$ = 14 meV, $T_C$ = 57 K for (b), and $\Delta_0$ = 16 meV, $T_C$ = 55 K for (d). (e) Symmetrized EDC at $k_F$ with temperature lower than 20 K for samples with different LTO thickness. Same figure as Figure 3 (c) in the main text.

# Section 7. Simulation of monolayer FeSe/LTO/STO

The simulation aims to calculate the supposed charge transfer amount for monolayer FeSe/LTO/STO heterostructure based on work-function balancing, without considering first-order phase transitions and insulating phase due to correlations. This simulation is based on an assumption that there is no hybridization between FeSe and LTO or STO electronic orbitals, which is partly justified by the ARPES measurement of monolayer FeSe/LTO/STO that shows basically identical band structure as doped multilayer/bulk FeSe. In particular, a tight-binding plus Poisson equation self-consistent calculation is used for the LTO/STO heterostructure to simulate the band bending and band structure. Electron correlation in LTO is not considered, thus the simulation is more accurate in the thin limit of LTO. This simulation is not intended to provide quantitative conclusion, but to clarify qualitative behaviors of the charge transfer across the FeSe and LTO/STO interface.

**List of notations**

$e$ : electron charge
$\varepsilon_0$ : electrostatic constant of vacuum.
$\varepsilon_r$ : effective dielectric constant between FeSe & oxide.
$n_t$: 2D density of transferred electron to FeSe.
$E = (n_t e)/\varepsilon_r \varepsilon_0$ : electric field between FeSe and oxide.
$V = E d$ : electric potential difference.
$d$ : distance between FeSe and oxide.
$W_{FeSe}(n_t)$ : FeSe work function as a function of $n_t$. $W_{LTO}(n_t)$ : LTO work function at the surface.
$c$ : coefficient for FeSe work function shift with doping.
$t$ & $t'$ : hopping energy for Ti 3d orbitals.
$k_x, k_y, k_z$ : electron momenta in LTO/STO.
$n_{2D}$ : total 2D electron density in all LTO/STO layers. $n_{La}$ : 2D electron density donated by La in one LTO layer. $l$ : number of LTO layers in the LTO/STO structure.
$r$ : the ratio of La atoms that donates electron.
$\varepsilon_{STO}$ : dielectric constant of STO.
$\varepsilon_{inf}$ : dielectric constant of STO at infinite electric field. $\varepsilon_{zero}$ : dielectric constant of STO at zero electric field.

**General description**

**Figure S7** schematizes the energy diagram for FeSe and LTO/STO heterostructure before and after contact. Before contact, since the additional LTO layers on top of STO serves as the electron donor to the system, electrons accumulate at the top surface and form a band bending, resulting in a lowered work function $W_{LTO}(n_t)$ (< 4.5 V, the work function of STO). When FeSe contacts with the LTO/STO, charge transfer occurs. Namely, LTO/STO loses electrons and FeSe gains electrons, which should make $W_{LTO}$ larger and $W_{FeSe}$ smaller. The quantitative amount of change of work function is associated with the amount of transferred electron $n_t$. With extra electrons, the FeSe monolayer is negatively charged. The created electric field builds a potential drop $V$ between the FeSe and the outmost oxide layer. As a consequence of the charge transfer, the chemical potential across the interface is balanced, giving rise to the following equation.

$$V(n_t) = W_{FeSe}(n_t) - W_{LTO}(n_t) \tag{1}$$

in which all three terms are functions of nt. By solving this equation, we can obtain the amount of transferred charge. In the following sections, each of these three terms will be discussed in details.

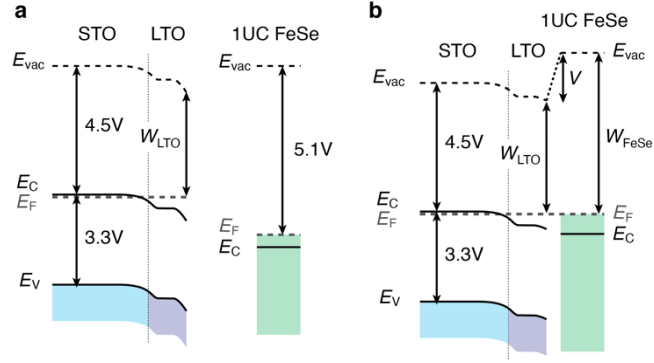

**Figure S7.** Schematics for FeSe and LTO/STO heterostructure chemical potential alignment before (a) and after (b) contact.

**Potential difference between FeSe and oxide $V(n_t)$**

It can be derived that $V(n_t) = n_t\, e\, d/(\varepsilon_r \varepsilon_0)$, in which $d = 8$Å is the spacing between FeSe and oxide, obtained from STEM literature,[10, 11] and $\varepsilon_r$ is the effective dielectric constant between FeSe and oxide. $\varepsilon_r$ is an important parameter, since it sets the capacitance value for the interface capacitor. We obtain this value based on experimental measurements of the FeSe/STO interface charge transfer.[12] Specifically, FeSe/STO has a charge transfer of $n_t \sim 0.11$ $e$/Fe; the STO band bending upward is measured to be 0.1 V; STO work function is 4.5 V; and the work function of FeSe on STO is measured to be 5.0 V. Thus, $V = 5.0$ V $-$ 4.5 V $-$ 0.1 V $= 0.4$ V, and $\varepsilon_r$ can be back calculated to be 52.

**LTO/STO surface work function $W_{\mathrm{LTO}}(n_t)$**

The accurate solution of the band bending at the surface and interface of this heterostructure requires a self- consistent method incorporating both band calculation and the spatially dependent electrostatics. In particular, the dielectric constant of STO is highly nonlinear and electric field dependent, spanning from more than $10^4$ to as low as $\sim 10$. Electric field forms a confinement potential quantum well at the surface of the LTO/STO heterostructure, and electrons in the 3d orbitals forms multiple subbands, or referred as quantum well states.

We integrate the Poisson equation for electrostatics and a tight-binding model for quantum mechanics to simulate the electron distribution of LTO/STO heterostructure. This model is adapted from reference.[13] Itinerant electrons occupy $d_{xy}$, $d_{yz}$, and $d_{xz}$ orbitals, in which electrons travel with a heavier mass in one direction and a lighter mass in the other two directions. The in-plane degree of freedom (spanned by $x$ and $y$ directions) has the simple solution shown below.

$$E_{\text{in-plane}} = \begin{pmatrix} -2t\cos k_x - 2t\cos k_y & 0 & 0 \\ 0 & -2t'\cos k_x - 2t\cos k_y & 0 \\ 0 & 0 & -2t\cos k_x - 2t'\cos k_y \end{pmatrix} \begin{Bmatrix} xy \\ yz \\ xz \end{Bmatrix}$$

in which $t$ and $t'$ are the hopping energy for electrons moving in the light and heavy directions, respectively. The values of these parameters are determined by fitting ARPES measured band structures. We find $t = 497$ meV, and $t' = 28$ meV.

In the out-of-plane $z$ direction, we solve it self-consistently considering both the tight-binding solution of electron distribution within the confinement potential quantum well and the specific shape of the quantum well which is calculated based on the electron distribution using Poisson equation (or Gauss's Law).

In particular, we look for eigenstates and eigenvalues of the sum of the hopping matrix and the electric potential matrix.

$$H_{dxy} + U = \begin{pmatrix} 0 & t' & 0 & 0 & \cdots \\ t' & 0 & t' & 0 & \cdots \\ 0 & t' & 0 & t' & \cdots \\ 0 & 0 & t' & 0 & \cdots \\ \vdots & \vdots & \vdots & \vdots & \ddots \end{pmatrix} + \begin{pmatrix} u_1 & 0 & 0 & 0 & \cdots \\ 0 & u_2 & 0 & 0 & \cdots \\ 0 & 0 & u_3 & 0 & \cdots \\ 0 & 0 & 0 & u_4 & \cdots \\ \vdots & \vdots & \vdots & \vdots & \ddots \end{pmatrix}$$

$$H_{dyz} + U = \begin{pmatrix} 0 & t & 0 & 0 & \cdots \\ t & 0 & t & 0 & \cdots \\ 0 & t & 0 & t & \cdots \\ 0 & 0 & t & 0 & \cdots \\ \vdots & \vdots & \vdots & \vdots & \ddots \end{pmatrix} + \begin{pmatrix} u_1 & 0 & 0 & 0 & \cdots \\ 0 & u_2 & 0 & 0 & \cdots \\ 0 & 0 & u_3 & 0 & \cdots \\ 0 & 0 & 0 & u_4 & \cdots \\ \vdots & \vdots & \vdots & \vdots & \ddots \end{pmatrix}$$

$$H_{dxz} + U = \begin{pmatrix} 0 & t & 0 & 0 & \cdots \\ t & 0 & t & 0 & \cdots \\ 0 & t & 0 & t & \cdots \\ 0 & 0 & t & 0 & \cdots \\ \vdots & \vdots & \vdots & \vdots & \ddots \end{pmatrix} + \begin{pmatrix} u_1 & 0 & 0 & 0 & \cdots \\ 0 & u_2 & 0 & 0 & \cdots \\ 0 & 0 & u_3 & 0 & \cdots \\ 0 & 0 & 0 & u_4 & \cdots \\ \vdots & \vdots & \vdots & \vdots & \ddots \end{pmatrix}$$

in which $u_i$ is the electric potential at the $i$th unit cell counting from the outmost surface of the LTO/STO heterostructure. With a specific known quantum well (i.e. known $u_i$), we are able to obtain the eigenstates and eigenvalues. Considering electric potential energy, kinetic energies from all three $x$, $y$, and $z$ directions, $E_F$ is determined by the total 2D electron density $n_{2D}$ (a function of $E_F$). $n_{2D} = l\, n_{La} - n_t$, in which $l$ is the number of total LTO layers, and $n_{La}$ is the electron density donated by La in one single LTO layer. $n_{La} = r \times 6.6 \times 10^{18}$, in which $r$ is the ratio of La atoms that actually donate electrons. With known eigenstates, eigenenergies and $E_F$, we can further calculate the electron distribution in each unit cell in the LTO/STO heterostructure. On the other hand, with the known electron distribution, we calculate the electric potential (i.e. $u_i$) based on Poisson equation incorporating the electric field dependent dielectric constant of STO $\varepsilon_{STO} = \varepsilon_{inf} + \varepsilon_{zero}/(1+E/E_C)$, in which $\varepsilon_{STO}$ is the dielectric constant function, $\varepsilon_{inf}$ is the dielectric constant when electric field is infinite, $\varepsilon_{zero}$ is the dielectric constant at zero field, $E$ is the electric field, and $E_C$ is a constant. We fit the 10 K data from ref.[14] to obtain the constants: $\varepsilon_{zero} = 20700$, and $E_C = 1 \times 10^5$ V/m.

**Results and Discussions**

We solve the problem $V(n_t) = W_{FeSe}(n_t) - W_{LTO}(n_t)$. **Figure S8** shows the simulation results of LTO/STO electron distribution and band bending before and after contact with FeSe, in which Figure S8a & S8e are results based on reference[12] for comparison. As can be seen, for all thicknesses of LTO, the band bending is significantly decreased after charge transfer, but the accumulated electrons at the interface is not depleted. We also find that the transferred electrons are mostly originally located at the first two unit cells before contact. **Figure S9** shows the amount of transferred charge to FeSe $n_t$ simulated in the model as a function of LTO thickness. As can be seen, $n_t$ surges rapidly for the first and second LTO layers, then mostly saturates thereafter, with an ultimate doping that is about twice as large as the 0UC (bare STO) case.

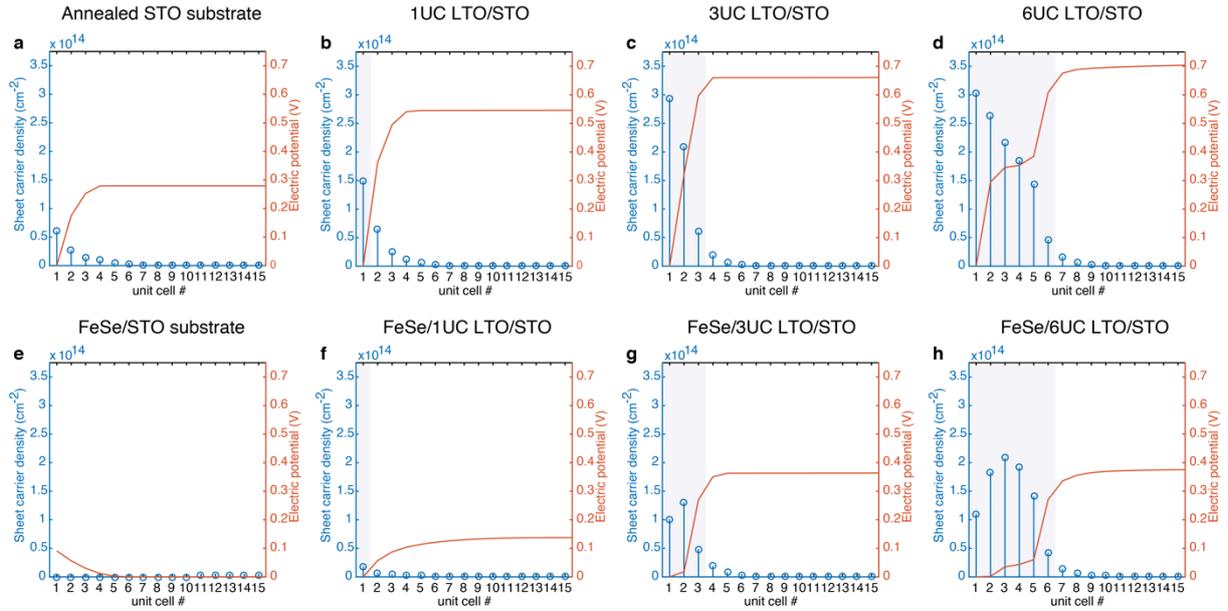

**Figure S8**. Electron distribution and band bending in the LTO/STO heterostructure simulated with the described model. (a) & (e) are results based on reference.[12] (a-d) are cases before contact with FeSe. (e-h) are cases after contact with FeSe. Blue shaded regions indicate the position of LTO layers.

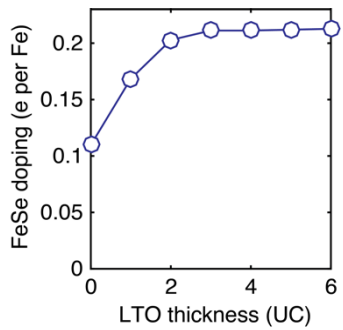

**Figure S9**. transferred charge to FeSe $n_t$ simulated in the model as a function of LTO thickness.

**Supporting Information References**